\documentclass[pdflatex,sn-mathphys-num]{sn-jnl}
\usepackage{graphicx}%
\usepackage{multirow}%
\usepackage{amsmath,amssymb,amsfonts}%
\usepackage{amsthm}%
\usepackage{mathrsfs}%
\usepackage[title]{appendix}%
\usepackage{xcolor}%
\usepackage{textcomp}%
\usepackage{manyfoot}%
\usepackage{booktabs}%
\usepackage{algorithm}%
\usepackage{algorithmicx}%
\usepackage{algpseudocode}%
\usepackage{listings}%

\begin{document}

\title[Article Title]{Spin quantum computing, spin quantum cognition}

\author*[1,2,3]{\fnm{Betony} \sur{Adams}}\email{betony@sun.ac.za}

\author[1,2]{\fnm{Francesco} \sur{Petruccione}}

\affil*[1]{School of Data Science and Computational Thinking and Department of Physics, Stellenbosch University, South Africa}

\affil[2]{National Institute for Theoretical and Computational Sciences, South Africa}

\affil[3]{The Guy Foundation, Dorset, UK}


\abstract{Over two decades ago, Bruce Kane proposed that spin-half phosphorus nuclei embedded in a spin-zero silicon substrate could serve as a viable platform for spin-based quantum computing. These nuclear spins exhibit remarkably long coherence times, making them ideal candidates for qubits.  Despite this advantage, practical realisation of spin quantum computing remains a challenge. More recently, physicist Matthew Fisher proposed a hypothesis linking nuclear spin dynamics, specifically those of phosphorus nuclei within the spin-zero matrix of calcium phosphate molecules, to neural activation and, potentially, cognition. The theory has generated both interest and scepticism, with some fundamental questions remaining. We review this intersection of quantum computing and quantum biology by outlining the similarities between these models of quantum computing and quantum cognition. We then address some of the open questions and the lessons that might be learned in each context. In doing so, we highlight a promising bidirectional exchange: not only might quantum computing offer tools for understanding quantum biology, but biological models may also inspire novel strategies for quantum information processing.}

\keywords{Kane quantum computer, nuclear spin, Posner molecules, neural activation}



\maketitle

\section{Introduction}

Both quantum computing and quantum biology focus on novel applications of quantum theory. Quantum computing is conventionally referred to as having taken shape in the 1980s. While a number of key thinkers contributed to the discussion \cite{feynman2}, Richard Feynman is often singled out as arguing that quantum rather than classical computers would be necessary to efficiently simulate quantum systems \cite{feynman1982,feynman2}. Early theoretical breakthroughs, such as Shor’s algorithm for factoring large numbers and Grover’s search algorithm, demonstrated the transformative potential of quantum computing \cite{shor1994,grover1996,grover1997}. On the technological side of things, the aim has been turning this theory into scalable machines. This remains a profound challenge, largely due to the extreme sensitivity of quantum states to environmental noise, which leads to decoherence and error accumulation.\\
\\
At the heart of this effort is the qubit -- the quantum bit -- which can exist in a superposition of states, enabling powerful computations. Many physical implementations have been pursued, including superconducting qubits, trapped-ion qubits, photonic qubits, and topological qubits. Silicon-based spin qubits are another promising contender, in particular for their long coherence times and compatibility with existing semiconductor technology \cite{ladd,jazaeri,chae,deleon,oukaira}. An early model for spin-based computing is the Kane quantum computer, proposed by Bruce Kane in 1998 \cite{kane}. It uses the nuclear spins of phosphorus-31 atoms embedded in silicon as qubits, taking advantage of their exceptionally long coherence times. Control is achieved through the use of radiofrequency magnetic fields and electrode gates, to manipulate the spins of each phosphorus nucleus and its bound electron \cite{kane,simmons2003,dzurak1}.\\
\\
The history of quantum biology goes even further back. When Feynman famously exclaimed that ``Nature isn’t classical, dammit'' \cite{feynman2}, he echoed the founding figures of quantum mechanics, who had raised the possibility that quantum theory might offer some insights into understanding living systems. This speculation culminated most explicitly in Schrodinger's 1944 book `What is Life?', which has been credited with inspiring the discovery of DNA \cite{mcfadden,marais,bohr,schrodinger}. Progress in quantum biology, however, has since been hindered by similar issues to quantum computing, namely decoherence. Indeed, it is a well-worn cliche that biological systems are too warm, wet, and messy to sustain quantum effects on any biologically meaningful timescales.\\
\\
Despite this scepticism, quantum biology is an expanding field of research, with biological, physiological, and medical relevance. Defined very simply, quantum biology looks at whether novel quantum effects such as superposition, coherence, tunnelling, and entanglement play a functionally relevant role in biological systems. For instance, quantum coherence in energy and charge transfer has been hypothesised to facilitate photosynthesis, where excitonic or vibronic coherence may enhance the efficiency of light-harvesting, as well as in microtubule dynamics and signalling processes \cite{engel,brixner,vangrondelle,schlau,panit,collini,craddock14,craddockbeats,kurian,celardo}. A further cornerstone of quantum biology is quantum tunnelling, historically observed in enzymatic reactions, where proton or electron tunnelling significantly accelerates reaction rates compared to classical expectations \cite{devault,klinman}. This mechanism has been extended to DNA mutation \cite{lowdin,slocombe} as well as receptor biology, for example, olfactory receptors have been proposed to operate via vibration-assisted tunnelling \cite{turin96,turinVTO}. Similar tunnelling-based models have been applied to neurotransmitter function, including serotonin \cite{hoehn}, as well as spike protein binding in coronaviruses \cite{adamscovid}. \\
\\
Another major area of study is the radical pair mechanism, an example of spin chemistry first developed to explain avian magnetoreception, in which the spin states of two correlated electrons respond to the Earth’s magnetic field, enabling birds to navigate \cite{schulten78,mouritsen2005,ritz2000,horerodgers}. Research is now expanding this framework to other biological processes, particularly the role of radical pairs in the generation of reactive oxygen species (ROS), which are central to inflammation and redox signalling \cite{simon1,simon2,simon3,simon4,usselman,usselman2}. Beyond electron spins, there has also been the suggestion that entangled phosphorus nuclear spin in calcium phosphate molecules known as Posner clusters might influence brain function and cognition \cite{fisher,fisher2,fisher3}. While experimental verification of this theory is still in progress, it has generated some interest from the point of view of quantum information theory \cite{halpern}. What is interesting about this particular theory of quantum cognition is its similarity, in physical implementation, to spin-based model architectures of quantum computing, particularly the Kane quantum computer.\\
\\
In this review article we compare and contrast these two models of spin-based information processing -- the Kane quantum computer and the Posner model of quantum cognition -- before outlining some of the questions that remain and the mutual insights each may offer the other. Human technology has a long history of biomimicry, drawing inspiration from the ingenuity of biological engineering. If living systems have indeed succeeded in harnessing functional quantum effects at room temperature, then quantum biology may serve as inspiration for the implementation of robust quantum computers. On the other hand, quantum neurobiology -- where the focus is on nerves and brains -- is still an underdeveloped aspect of the broader field of quantum biology, and might borrow impetus from the intensive research being done in the context of quantum information processing and quantum computing.

\begin{figure}[h]
\centering
\includegraphics[width=0.9\textwidth]{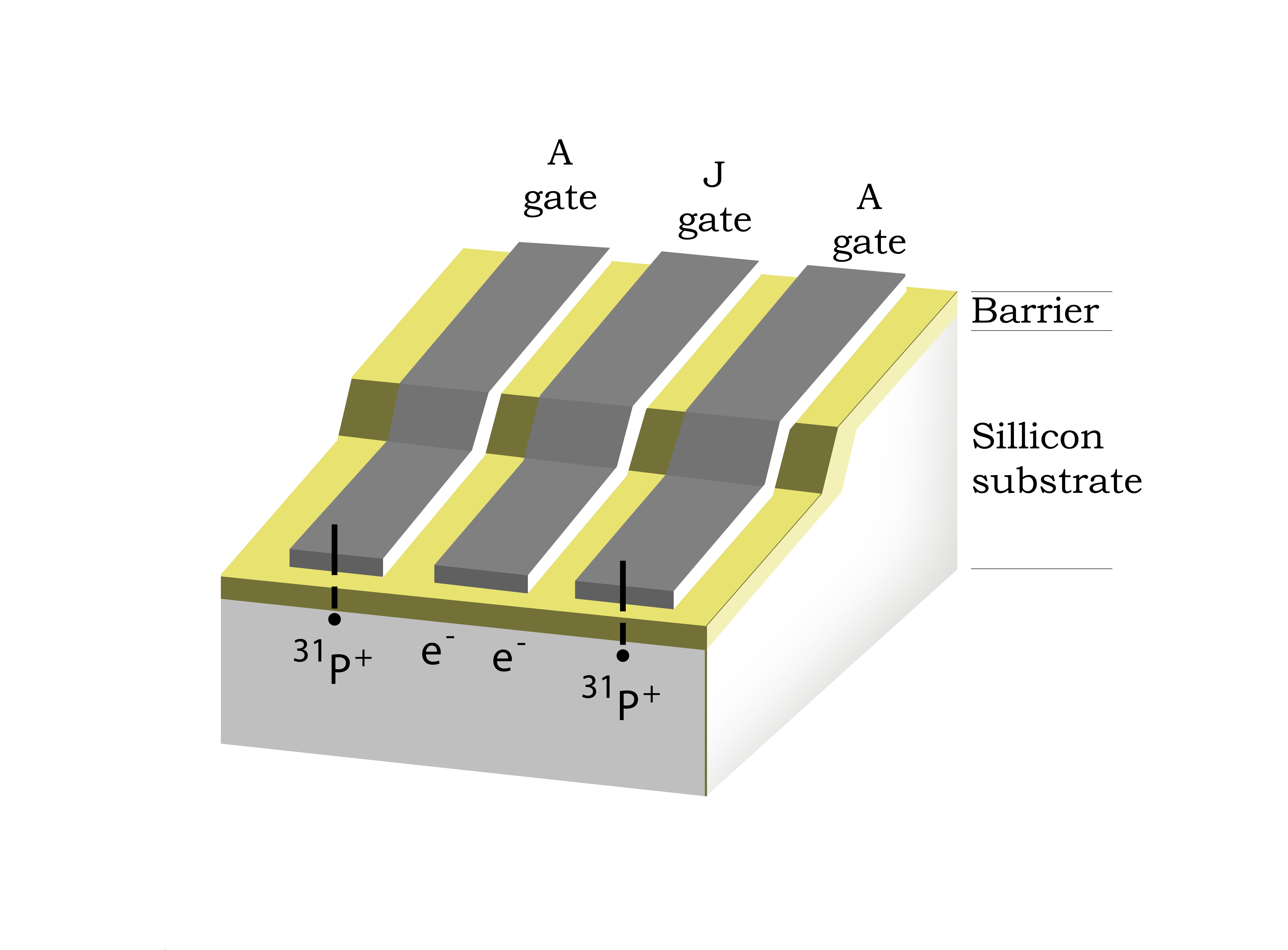}
\caption{Simple schematic diagram of a nuclear-spin-based quantum computer. The Kane quantum computer employs phosphorus-31 atoms embedded in isotopically pure silicon, where each donor nucleus serves as a qubit. Each phosphorus atom has a weakly bound electron whose spin couples to the nuclear spin via the hyperfine interaction. Control is achieved using surface electrodes: A-gates tune the hyperfine coupling between a donor nucleus and its electron, while J-gates adjust the overlap of neighbouring electrons to mediate interactions between qubits. Readout relies on transferring nuclear spin information to the donor electron and detecting resulting charge shifts with a single-electron transistor (not shown), exploiting the Pauli exclusion principle to distinguish singlet and triplet states \cite{kane,simmons2003,dzurak1}.}
\end{figure}

\section{Spin quantum computing}
\subsection{The qubit}
In quantum computing, the qubit is the fundamental unit of computation, analogous to the classical bit but with uniquely quantum properties, such as superposition of states, that allow greater computational potential. The simplest physical implementation of a qubit is a quantum two-level system, in which the logical states are represented by two well-defined quantum states of a physical system. For example, the ground and excited states of an atom, the polarisation states of a photon, or the two possible spin orientations of an electron, can each serve as the basis of a qubit \cite{divincenzo,ladd,jazaeri,chae,deleon,oukaira}. This has led to the investigation and development of different quantum computing architectures, exploiting different quantum systems. In this paper we focus only on those models of quantum computing that use spin degrees of freedom as the physical implementation of a qubit.\\
\\
Spin-based quantum computing includes a number of different possible architectures. The qubits in question might also refer to either electron or nuclear spin. For the purposes of this paper we focus on a specific iteration of spin quantum computing that developed from a model first proposed by Bruce Kane, in which the nuclear spin of phosphorus-31 atoms functions as the qubit of choice \cite{kane}. The Kane quantum computer constitutes a scalable, repeating unit cell that resembles the architecture of modern semiconductor devices. It is built from a silicon crystal doped with phosphorus atoms which are controlled by surface control electrodes, termed the A and J gates. In addition, the phosphorus dopant has a bound electron, whose spin interacts with phosphorus nuclear spin through the hyperfine effect. This spin interaction allows for additional control of the nuclear spin qubit
\cite{kane, dzurak1,simmons2003}. While the exact manifestation of the spin qubit has undergone some evolution since Kane's first concept paper -- including the introduction of the `flip-flop' qubit -- silicon-based quantum computing is still generating interest, in part due to its compatibility with existing computing technology \cite{savytskyy,reiner,pla,tosi}.

\subsection{Protecting, entangling, and reading the qubit}
Of all qubit contenders, nuclear spin is optimal from the point of view of decoherence, with reported coherence lifetimes of over half an hour \cite{zwanenburg,hill,saeedi}. The long coherence times of nuclear spin qubits in the Kane quantum computing architecture depend on the accurate placement of phosphorus atoms in a perfect spin-zero, isotopically pure silicon-28 crystal, a fabrication challenge that has come some way since Kane's original hypothesis \cite{simmons2003,pla,weber,acharya}. However, while nuclear spin qubits are promising in terms of coherence lifetimes, this isolation from interaction is in part a double-edged sword, making qubit manipulation more challenging. \\
\\
In early models of the Kane quantum computer, qubit operations were achieved through magnetic-field manipulation of nuclear spin. In an applied magnetic field the degeneracy of the nuclear spin-up and spin-down states is lifted and spins oriented parallel to the field occupy a lower energy state than those anti-parallel. By applying a radiofrequency pulse with the appropriate frequency, the nuclear spin can be driven between the lower and higher  energy state, effectively flipping its orientation \cite{kane,simmons2003,dzurak1}. For more than one nuclear spin, these radiofrequency pulses can be tailored to pick out a specific nucleus by manipulating the hyperfine interaction between the phosphorus nucleus and its bound electron. This hyperfine tuning is achieved through the A-gate electrodes. The J-gates, on the other hand, are used to manipulate the interaction between nuclear spin qubits. This is achieved by indirect spin coupling through intermediary electron spins, where the J-gates alter the overlap of neighbouring electrons \cite{kane,simmons2003,dzurak1}.\\
\\
Since nuclear spins do not interact strongly with one another over the relevant distances in silicon, their entanglement must also be mediated by their coupled electrons. For example, in the original Kane computer architecture, tuning the J-gate means two phosphorus donor electrons can be forced to overlap and the exchange interaction can be exploited to entangle their spin states. Manipulation of the exchange interaction (between electrons) and the hyperfine interaction (between electron and nucleus) allows the quantum correlations of the electrons to be transferred to the nuclear spins \cite{stemp,stemp2}. In this way, the Kane architecture achieves nuclear spin entanglement indirectly, using the electrons as intermediaries to link otherwise isolated phosphorus nuclei. The exchange interaction poses a fabrication challenge, due to its short range of influence. Nonetheless, progress has been made in demonstrating the experimental implementation of exchange-based entanglement in electrons bound to individual phosphorus donors in silicon \cite{stemp,stemp2}.\\
\\
Finally, readout of computations is achieved by converting spin detection into charge detection, the latter of which can be measured by a suitably positioned single electron transistor device (SET) \cite{kane,simmons2003,dzurak1}. Once again this is an indirect readout of qubit spin state, achieved through the transfer of information from nuclear to electron spin. By manipulating electron-nuclear and electron-electron states through A and J-gates, two-electron states can be switched between triplet and singlet spin states. The Pauli exclusion principle then allows for the differentiation between these two states, given that in the event of an electron transfer between donors only the singlet state forms a stable two-electron bound state on a phosphorus atom \cite{kane,simmons2003,dzurak1}.\\
\\

\section{Spin quantum cognition}

\begin{figure}[h]
\centering
\includegraphics[width=1\textwidth]{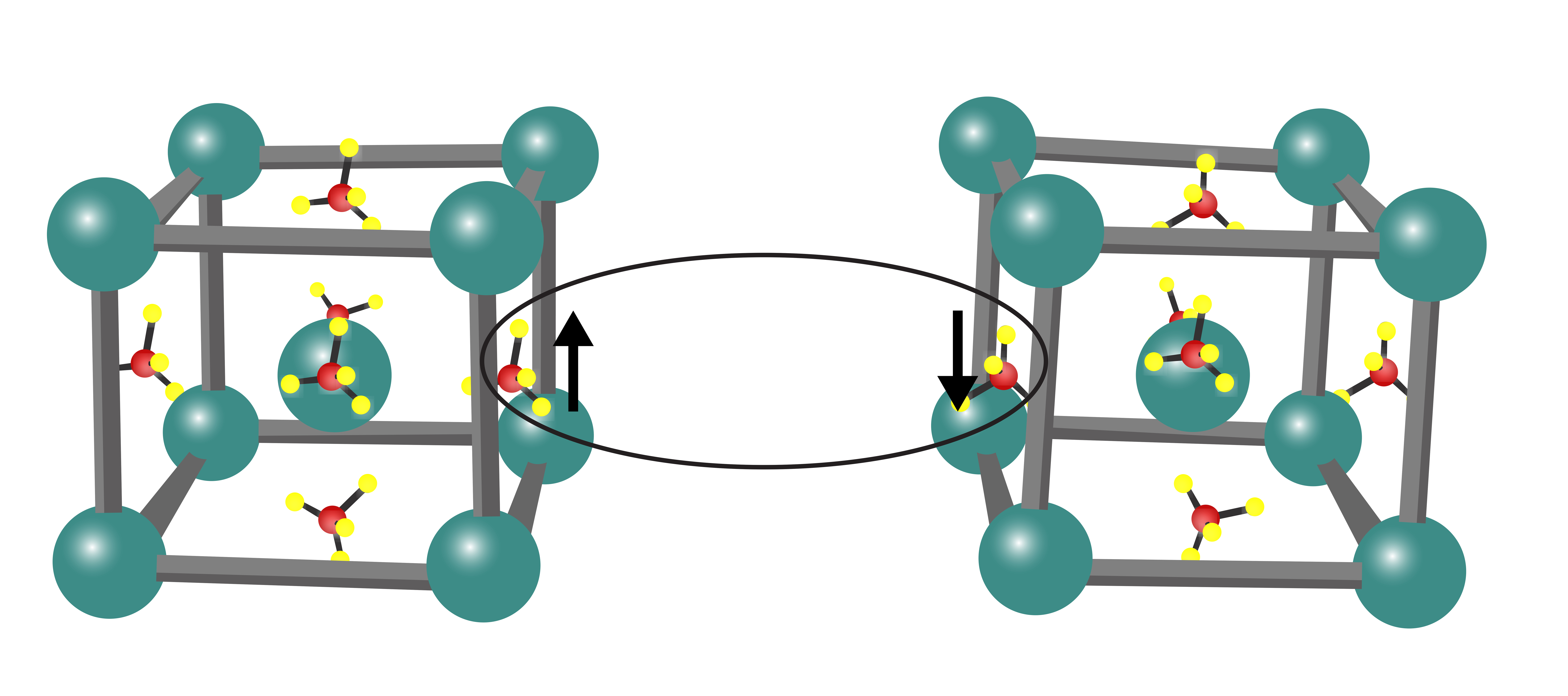}
\caption{Simplified illustration of entangled Posner molecules. Each Posner molecule consists of nine calcium ions  (blue spheres), eight of these on the corners of a cube, with one in the centre of the cube. On the faces of the cube are the six phosphates, with each phosphate consisting of one phosphorus (red) and four oxygen (yellow) ions. Entanglement between Posner molecules is conferred by the entanglement of phosphorus nuclear spins in different Posner molecules. In this illustration the black circle and arrows shows one entangled phosphorus pair in a singlet state (antiparallel spins).}
\end{figure}
\subsection{The qubit}
In quantum computing a number of different candidates for physical qubits have been explored. Similarly, the biological equivalent of a `qubit' might refer to various different quantum systems.  In well-studied cases such as photosynthesis, it has been suggested that energy transfer along pigment–protein complexes might employ quantum coherence, where electronic excitations migrate with remarkable efficiency across chromophores. These quantum effects operate on ultrafast timescales -- femtoseconds to nanoseconds \cite{engel,brixner,vangrondelle,schlau,panit,collini}. However, for quantum effects to play a role in cognition, timescales must reach at least milliseconds or longer, to have any relevance for neural signalling. In the field of quantum biology, longer-lived quantum states involve the spin of electrons and nuclei. For example, the radical pair model of avian magnetoreception illustrates how spin-sensitive reactions can couple quantum states to biochemical signalling \cite{schulten78,mouritsen2005,ritz2000,horerodgers}. Radical pairs involve two electrons that start out in a specific quantum state: either a singlet (antiparallel) or a triplet (parallel spins). External magnetic fields, and local nuclear spins, can cause the electron spins to oscillate between singlet and triplet states. Since the singlet and triplet states often lead to different chemical products, the reaction yield becomes sensitive to magnetic interactions. Radical pairs have been suggested to have lifetimes in the microsecond range \cite{schulten78,mouritsen2005,ritz2000,horerodgers}.\\
\\
While electron spin is a promising candidate for longer-lived quantum states in biological systems, decoherence remains an issue. For quantum cognition, more stable qubits are necessary. For this reason, physicist Matthew Fisher suggested that nuclear spin is a better contender for spin quantum cognition \cite{fisher,fisher2,fisher3}. In particular spin-half nuclei have very long coherence times, due to the fact that they have no quadrupole interactions. Common elements with spin-half nuclei in biological systems are limited to hydrogen and phosphorus. This led to the hypothesis that entangled phosphorus nuclear spin in calcium phosphate molecules might play a role in quantum cognition \cite{fisher,fisher2,fisher3}. 

\subsection{Protecting, entangling, and reading the qubit}
While phosphorus nuclear spin does not experience the faster decohering effects of the quadrupole interaction, it still interacts with surrounding nuclear spins \cite{fisher,fisher2,fisher3}. In particular, biological systems, given their high water content, contain an abundance of hydrogen, which has non-zero nuclear spin. In order to prevent decoherence it is suggested that phosphorus nuclear `qubits' are protected by the fact that they bind with oxygen into phosphates, as well as with calcium, to create a specific form of calcium phosphate known as a Posner molecule \cite{fisher,fisher2,fisher3,posner}. Calcium phosphate is extremely prevalent in the body, being the main constituent of bone but also an important reservoir for calcium ions, which in turn are very important in the activation of nerves \cite{wolf,sudhof,neher}. As both oxygen and calcium have zero nuclear spin, this results in a spin-zero substrate for the phosphorus nuclei, in a manner similar to the Kane quantum computer's use of purified silicon. This led Fisher to postulate that phosphorus nuclei could have unprecedented coherence lifetimes, up to 21 days \cite{fisher,fisher2,fisher3}. While this estimate generated scepticism, revised estimates of over half an hour seem equally impressive in the context of fragile quantum states \cite{player}. A more recent paper revisits this estimate and concludes that entanglement between two nuclear spins in separate Posner molecules decays on a sub-second timescale \cite{agarwal1,agarwal2}.\\
\\
In the Kane quantum computer, nuclear qubits can be manipulated through the judicious use of applied fields. In a similar manner, phosphorus nuclear spin in biological substrates might be manipulated by local magnetic interactions. While ideal Posner molecules consist of a spin-zero `shell' of calcium and oxygen atoms, in reality calcium phosphate molecules can incorporate a number of other trace biological elements \cite{mancardi}. Indeed, it has been suggested that the substitution of lithium for the central calcium in Posner molecules may account for the different effects that lithium isotopes have on mothering behaviour in rodents \cite{fisher,ettenberg,sechzer}. Lithium is a powerful psychiatric medicine, used primarily in the treatment of bipolar disease. Lithium-6 and lithium-7 have different nuclear spin -- of $1$ and $\frac{3}{2}$ respectively -- which would mean that each isotope would have a different spin interaction with the phosphorus nuclear spin in a lithium-doped Posner molecule \cite{fisher,adams2025}. This differential interaction could in turn have a different effect on free calcium ion concentration and neural excitability, which appears to play some role in bipolar disease \cite{stern,mertens}. In this way, the different nuclear spin states of lithium isotopes translate into different pharmacological outcomes. Recent experimental evidence supports this hypothesis by demonstrating that lithium isotopes differentially influence calcium phosphate mineralisation in biologically relevant aqueous solutions \cite{patel}.\\
\\
Entanglement is one of the primary novel quantum resources by which quantum computing might outperform its classical counterpart. In the Kane quantum computer, entanglement of nuclear qubits is achieved indirectly, through the interaction with entangled electrons. In the quantum cognition case, Fisher's original hypothesis outlined how enzyme-facilitated nuclear spin entanglement might occur chemically \cite{fisher,fisher2,fisher3}. In biology, pyrophosphate is generated as a by-product of adenosine triphosphate (ATP), the primary energy currency of the cell. Fisher originally proposed that when pyrophosphate undergoes enzymatic cleavage, it yields two phosphate ions whose phosphorus nuclei are released in a singlet spin state, which is an entangled state \cite{fisher}. According to his hypothesis, the enzyme pyrophosphatase constrains the rotational freedom of the pyrophosphate molecule during hydrolysis, and the symmetry requirements for identical fermions (phosphorus spin-half nuclei) make the singlet configuration the favourable outcome \cite{fisher}.\\
\\
Entanglement in this instance thus refers to the singlet state in which the correlated nuclear spins begin. Entanglement in radical pair reactions is often invoked in a similar manner, although electron entanglement is not necessary for spin-dependent chemistry to take place, which relies on spin-selective reactivity \cite{hogben,gauger,kattnig}. In Fisher's hypothesis, the individual phosphorus nuclear spins are then subsequently described in terms of pseudospin, a property of the molecule as a whole which follows from the particular symmetry of Posner molecules. The interaction of pseudospin with the rotational states of the molecule then constrains molecule interactions, effectively imposing spin conditions on chemical binding outcomes \cite{fisher}. However, more recent papers have raised some questions about spin correlation and entanglement in Posner molecules, in particular whether Posner molecules naturally exhibit the required symmetry \cite{agarwal1}. In terms of entanglement transfer between individual phosphorus nuclei within a Posner molecule, symmetry also appears to be an important feature, with asymmetric configurations yielding low entanglement transfer in spin simulations \cite{player,agarwal1,agarwal2,adams2025}. Posner molecules, with a total of six phosphorus spins, are also not optimal for preserving entanglement, due to intramolecular dipolar interactions, with some researchers suggesting that calcium phosphate dimers, with only four phosphorus nuclei are better suited for this purpose \cite{agarwal2}.\\
\\
But how does this entanglement result in a change in cognitive state? In quantum computing, phosphorus nuclear qubits are used as a way to encode information. In quantum cognition, phosphorus nuclei mediate quantum effects in nerves, which have a knock-on effect on cognition. The mechanism by which Fisher suggests this happens is the following: if entangled phosphorous nuclei in amorphous calcium phosphate clusters known as Posner molecules are taken up by neurons, their quantum state might influence the release of calcium ions during synaptic firing \cite{fisher,fisher2,fisher3}. Since calcium release is a key trigger for neurotransmitter vesicle release, the argument is that entangled nuclear spins could bias the probability of neurotransmission in a correlated, non-classical way. In this way, nuclear spin entanglement in Posner molecules could affect nerve signalling and thus, potentially, cognition. How does `readout' of nuclear spin states occur in this biological context? In the Kane quantum computer, nuclear spin state is conventionally read out indirectly, through its interaction with a coupled electron \cite{kane,simmons2003,dzurak1}. In the Posner model of quantum cognition, nuclear spin states are coupled to molecular rotational states, which in turn influence molecular binding probability and subsequent hydrolysis and free ion concentration. Molecular binding, in this sense, acts as a readout of nuclear spin state. \\
\\

\section{Open questions and mutual insights}
Although the Kane quantum computer and the Posner molecule model of quantum cognition arise from very different contexts -- engineered silicon devices versus biological chemistry -- they share some striking similarities in their design principles. In both cases, phosphorus nuclei serve as long-lived qubits, protected from decoherence by their spin-zero local environments and coupled through electron-mediated interactions. In the Kane computing architecture, these electron intermediates are manipulated using J-gates which target the exchange interaction, and A-gates which target the hyperfine interaction \cite{kane,simmons2003,dzurak1}. In the Posner model, the indirect spin coupling, or J-coupling, is the primary way in which the phosphorus nuclear spins interact. Indirect spin coupling is the interaction of nuclear spin through intermediate electrons, and the strength of this interaction can have profound effects on quantum parameters such as coherence and entanglement \cite{player,agarwal1,agarwal2,adams2025}.\\
\\
Given these similarities, we were intrigued to consider whether progress made in understanding the strengths and limitations of each of these contexts might have insights for the other. For example, the Posner molecule model to some extent lacks a convincing mechanism for quantum to biological transduction, or -- in the computing sense -- a readout mechanism. Fisher's hypothesis that pseudospin and rotational entanglement leads to altered chemical binding appears to depend on Posner molecule symmetry \cite{fisher,fisher2,fisher3}, a property that has been thrown into some doubt \cite{agarwal1}. In the radical pair mechanism, which describes the interaction of electron spins, quantum to biological readout is well described due to the fact that the Pauli exclusion principle translates spin states into biochemical products \cite{schulten78,mouritsen2005,ritz2000,horerodgers}. It is less well understood how nuclear spin states might translate into functional chemical or biological outcomes.\\
\\
The efficacy of nuclear spin as a qubit demands a tradeoff: isolation ensures long coherence times but complicates accessibility to the quantum state and subsequent readout. Spin quantum cognition might borrow from techniques employed in spin quantum computing, namely the transferral of spin information from nuclei to electrons, which are then used as a means for reading out the quantum state. In this way nuclear spin in protected substrates such as Posner molecules might act as a `storage' of quantum information, then at some later stage this information is transferred to electrons which then proceed to a biochemical signature via mechanisms such as the radical pair and spin chemistry. This would likely require some elaboration on the conventional Posner model, namely the presence of an electron spin interacting with the phosphorus nuclear spin. While reactive oxygen species have generated a great deal of research interest, due to their central role in signalling and inflammation processes, phosphates can also form radical species, though their role in biological processes is not as well understood \cite{buhl,ma,criado}.\\
\\
Interestingly, in the context of quantum biology, there has also been some discussion of the role of radical pairs in phosphorylation processes related to N-methyl-D-aspartate (NMDA) receptors \cite{simon5}. NMDA receptors are crucial to brain function and several experiments have demonstrated that they exhibit sensitivity to static and oscillating magnetic fields \cite{simon5,ozgun,salunke}. It is suggested that this magnetic sensitivity is related to phosphorylation, citing previous research that demonstrates a spin-dependent reaction involving a phosphate oxyradical, formed from adenosine diphosphate (ADP) \cite{simon5,buchachenko1,buchachenko2}. Recent research into the quantum information processing abilities of ADP as a two-qubit system of phosphorus nuclear spin, showed coherence lifetimes of milliseconds, which is long enough to influence nerve firing \cite{Davino}. In the context of quantum models of cognition, phosphorylation is integral to the cognitive function of neurons, including memory formation \cite{carew,czernik,sharma}. It would thus be potentially interesting to reconsider the Posner molecule model of spin cognition from the point of view of phosphorylation, where longer-lived nuclear spin information might be translated into electron spin reaction dynamics.\\
\\
The nuclear-electron spin interaction that is central to qubit manipulation in the Kane computing architecture is an interesting lens through which to re-imagine quantum cognition. On the other hand, what insights might be leveraged in the opposite direction? Recent progress in phosphorus-based spin computing has focused on the scaling up of entanglement, where entanglement between nuclear qubits might be achieved through their interaction with coupled electrons \cite{stemp,stemp2}. Entanglement in the Posner model can either refer to Fisher's original conception of pseudospin, or more simply the relative alignment of two phosphorus nuclear spins, such as the entangled singlet state that is arguably produced by the hydrolysis of ATP before being incorporated into Posner molecules \cite{fisher,player,agarwal2,adams2025}. In the latter case, there has been some discussion of how different spin interactions, including the Zeeman and indirect spin coupling, convert this singlet to three possible triplet states \cite{player,agarwal2,adams2025}. While one of these triplet states is an entangled state, two are not entangled. Indirect spin coupling (J-coupling) strength, for example, has been demonstrated to have an effect on the spin dynamics and entanglement \cite{player,agarwal2,adams2025}. \\
\\
An interesting feature of the relative strengths of the spin interactions in Posner molecules is that J-coupling strengths are markedly weaker than the Zeeman interaction, which is the interaction of the nuclear spins with the geomagnetic field \cite{adams2025}. In radical pair research, a difference in relative strengths between the hyperfine and Zeeman interactions is known as the high field effect and results in the energy separation of the two non-entangled triplet states from the entangled singlet and triplet states. It has been suggested that this offers a way to create a subspace in which most of the mixed (singlet and triplet) states are entangled \cite{tiersch}. In the Posner context, this `high-field effect' is the natural outcome of the different scales of the J-coupling strengths with respect to the geomagnetic field \cite{adams2025}. The usefulness of this entanglement `concentration' is debatable in the biological context, especially given that Posner molecules are in diffusion and entanglement can be considerably degraded by the loss of position information \cite{eisert}. However, in the context of spin quantum computing, where the qubits are fixed, the high-field effect might offer a way to ensure that nuclear (or electron) spin is confined to entangled states.

\section{Conclusion}\label{sec13}

Spin-based models of quantum information processing, whether engineered in silicon or hypothesised to occur in biological matter, show an interesting convergence of design principles. There are also unresolved questions in both domains. For spin quantum computing, scalability and error correction remain significant challenges. For quantum cognition, experimental verification of long-lived nuclear spin coherence and functionally relevant entanglement is still outstanding. By comparing and contrasting these models, this paper is aimed at inspiring dialogue between quantum information science and quantum biology. If biological systems have indeed succeeded in harnessing nuclear spin as a resource for room-temperature quantum processing, they may offer blueprints for future quantum technologies and novel strategies for enhancing entanglement. Equally, advances in spin-based computing may help elaborate on hypotheses for quantum effects in cognition, for instance the long coherence times of nuclear spin in Posner molecules might be coupled to new readout mechanisms such as radical spin chemistry. Both fields might benefit from a shared understanding of how quantum principles can be stabilised, controlled, and translated into meaningful information processing, across silicon and synapse alike.

\backmatter

\bmhead{Acknowledgements}

B.A and F.P. were supported by the National Institute for Theoretical and Computational Sciences. Thank you to Angela Illing for the diagrams.


\bibliography{sn-bibliography}

\end{document}